\documentclass[twocolumn,showpacs,aps,prl,superscriptaddress,floatfix,amsmath,amssymb]{revtex4}

\usepackage{graphicx}
\usepackage{dcolumn}
\usepackage{bm}
\usepackage{subfigure}

\begin{document}

\title{Absence of broken time-reversal symmetry in the pseudogap state of the high temperature La$_{2-x}$Sr$_{x}$CuO$_{4}$
    superconductor from muon-spin-relaxation measurements}

\author{G. J. MacDougall}
\affiliation{Department of Physics and Astronomy, McMaster University,
Hamilton, ON, Canada, L8S-4M1}

\author{A. A. Aczel}
\affiliation{Department of Physics and Astronomy, McMaster University,
Hamilton, ON, Canada, L8S-4M1}

\author{J.P. Carlo}
\affiliation{Department of Physics, Columbia University,
New York, New York, USA, 10027}

\author{T. Ito}
\affiliation{Department of Physics, Columbia University,
New York, New York, USA, 10027}

\author{J. Rodriguez}
\affiliation{Department of Physics and Astronomy, McMaster University,
Hamilton, ON, Canada, L8S-4M1}

\author{P. L. Russo}
\affiliation{Department of Physics, Columbia University,
New York, New York, USA, 10027}
\affiliation{TRIUMF, Vancouver, British Columbia, Canada, V6T 2A3}

\author{Y. J. Uemura}
\affiliation{Department of Physics, Columbia University,
New York, New York, USA, 10027}

\author{S. Wakimoto}
\affiliation{Quantum Beam Science Directorate, Japan Atomic Energy Agency, Tokai, Ibaraki 319-1195, Japan}

\author{G. M.  Luke}
\affiliation{Department of Physics and Astronomy, McMaster University,
Hamilton, ON, Canada, L8S-4M1}
\affiliation{Canadian Institute for Advanced Research, Toronto, Ontario, Canada, M5G 1Z8}

\date{\today}
\begin{abstract}
We have performed zero-field $\mathrm{\mu SR}$ measurements on single crystals
of $\mathrm{La_{2-x}Sr_{x}CuO_{4}}$ to search for spontaneous currents in the
pseudo-gap state. By comparing measurements on materials across the phase diagram, we put strict upper limits on any possible time-reversal
symmetry breaking fields that could be associated with the
pseudo-gap. Comparison between experimental limits and the proposed circulating current states effectively eliminates the possibility that such states exist in this family of materials.
\end{abstract}


\maketitle

The nature of the pseudo-gap and its relation to superconductivity is among the largest outstanding problems in cuprate research. To date, many
theories of the pseudo-gap have been debated, but experiments
have not definitively elevated one above the rest\cite{timusk99,tallon01,norman05}.

One theory which has been receiving particular attention is the
`circulating current' picture of Varma\cite{varma,simon02}. In this proposal,
the pseudo-gap temperature, $\mathrm{T^{*}}$, represents a real phase transition
temperature to a novel ordered state, which is characterized by spontaneous currents
flowing in specific patterns in the copper-oxide planes. This line of phase
transitions terminates in a quantum critical point (QCP) inside the
superconducting dome, and the resulting quantum fluctuations are ostensibly
responsible for both the superconductivity and the non-Fermi-liquid-like
behavior at optimal doping.

Theoretical work on this proposal has been extensive, and the model reproduces many features of the cuprate phase diagram. However,
the experimental situation is much less conclusive, largely due to
the difficulty in coupling to the proposed order parameter. To date,
the main experimental case for this picture rests on two studies, both hotly contested.

\begin{figure}[tpb]
\centering
\subfigure{
\includegraphics[width=0.6\columnwidth]{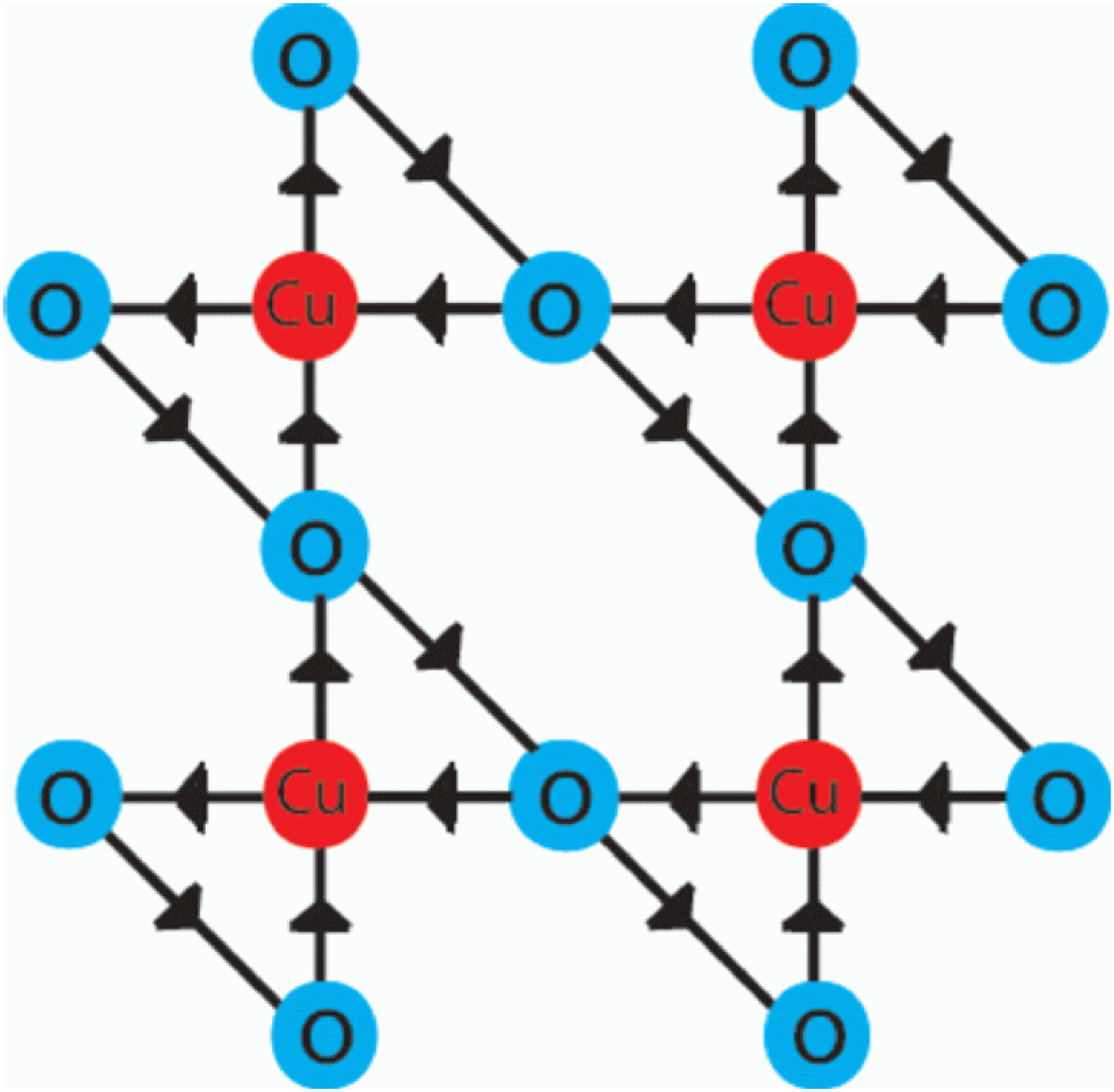}}
\subfigure{
\includegraphics[width=0.75\columnwidth]{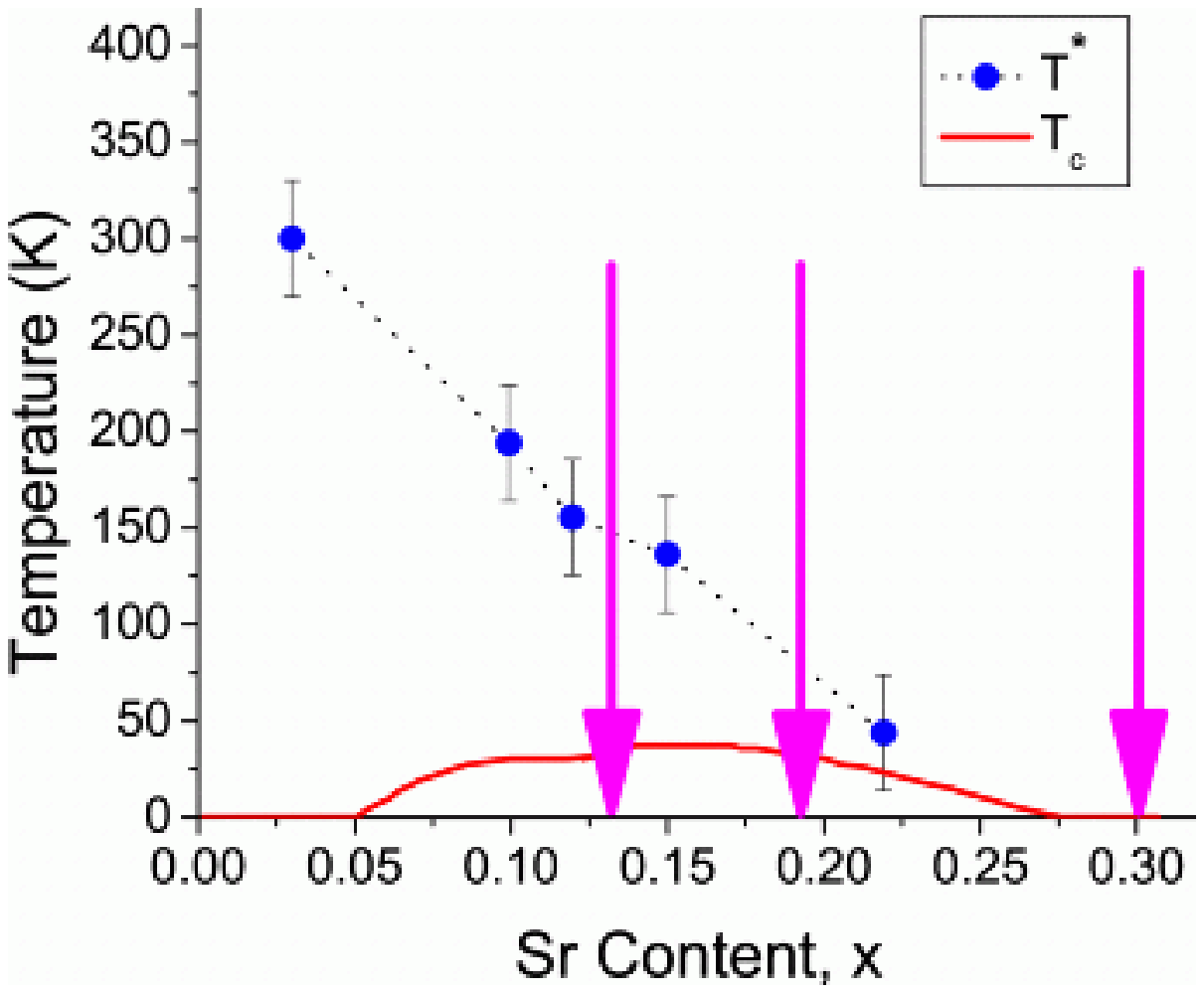}}
\caption{\label{fig:pattern}(Top)The orbital current pattern
  inferred from ARPES and neutron scattering data. (Bottom) The experimental phase diagram of $\mathrm{La_{2-x}Sr_{x}CuO_{4}}$. Data is taken from Ref.~\onlinecite{hashimoto07}. Arrows indicate the dopings and temperature range explored in the current study.}
\end{figure}

The first is the apparent observation of dichroism in the ARPES spectrum of
$\mathrm{Bi_{2}Sr_{2}CaCu_{2}O_{8+\delta}}$ (Bi2212)\cite{kaminski02}. Such dichroism is indicative of a
time reversal symmetry breaking (TRSB) field in the pseudo-gap state: a key prediction of the circulating current theory. Although the symmetry of the dichroism was not consistent with original predictions, an alternate current pattern was proposed which could explain the data\cite{simon02}. (See Fig.~\ref{fig:pattern}.)
However, there has been much debate about the importance of super-lattice structural distortions subsequently shown to exist in
this material\cite{castellan06}, and a later experiment on a material without these distortions
 saw no such effect\cite{borisenko04}.

More recently, a spin-polarized neutron scattering group reported
commensurate magnetic peaks below the pseudo-gap temperature in several
samples of $\mathrm{YBa_{2}Cu_{3}O_{6+x}}$ (YBCO)\cite{fauque06}. These results have since been reproduced in YBCO\cite{mook08} and also $\mathrm{HgBa_{2}CuO_{4+\delta}}$\cite{greven08}. The locations of the peaks in reciprocal space were consistent with the current pattern inferred from the ARPES data, and the authors of these papers clearly favor this interpretation. However, the authors have also been careful to note that relatively few magnetic peaks have been measured, and have given further
 examples of spin arrangements which could alternatively explain the data\cite{fauque06,sidis06}. Furthermore, the observed orientation of the local dipole moments was inconsistent with in-plane currents, and it was necessary to postulate a co-existing spin order to reconcile this fact with theory\cite{aji07}.

Given these concerns, it is evident that clarifying experiments are needed before drawing any general conclusions. Muon spin rotation ($\mathrm{\mu SR}$) is in an especially good position to illuminate the situation. It can readily detect the TRSB fields that would necessarily accompany the proposed current order. It probes local properties of the bulk, making it largely insensitive to impurity and surface effects which might confuse interpretation. In addition, its sensitivity to TRSB
 fields as small as 0.5~G  has been repeatedly demonstrated in other superconducting materials\cite{TRSB}
as well as in a long list of weak magnetic systems. The effect would be even more pronounced in the present scenario, as muons would precess coherently in the long-range ordered state, should it exist.

In an attempt to address the outstanding questions regarding circulating current order in the cuprates, we have performed zero-field $\mathrm{\mu SR}$ (ZF-$\mathrm{\mu SR}$) on several single crystals of $\mathrm{La_{2-x}Sr_{x}CuO_{4}}$ (LSCO), both in and outside of the pseudo-gap regime. LSCO was chosen because its pseudo-gap has been well mapped out\cite{hashimoto07}, and high quality single crystals exist with a wide range of carrier concentrations. This afforded us the ability to compare pseudo-gap and non-pseudo-gap materials directly by varying the amount of Sr in the system, and avoided the ambiguity involved in a defining the pseudo-gap temperature. In this way, we are able to effectively eliminate the possibility that the above circulating current state exists in LSCO.

Three crystals were grown via the traveling solvent floating zone method, with x=0.13 ($\mathrm{T^{*}\sim 150K}$), x=0.19 ($\mathrm{T^{*}\sim 75K}$) and x=0.30 ($\mathrm{T^{*}=0K}$), where the values of $\mathrm{T^{*}}$ are those reported by recent ARPES studies\cite{hashimoto07}. Growth conditions and sample properties are described in more detail elsewhere\cite{savici05,wakimoto04}. Samples were cut into $\sim$1mm thick plates with the large face perpendicular to the $\hat{c}$-axis. ZF-$\mathrm{\mu SR}$ spectra were taken in a He-flow cryostat on the M20 beamline at TRIUMF, in Vancouver, Canada. This allowed us to explore the region of the phase diagram indicated by the arrows in the bottom panel of Fig.~\ref{fig:pattern}, crossing the pseudo-gap transition line along both the temperature and doping axes.

Muons were implanted one at a time, with their spins initially polarized $\sim$45$^{\circ}$ from the beam axis ($\| \hat{c}$), and allowed to evolve in the presence of the local internal field. Muons decay with a mean lifetime
 $\tau_\mu =2.2\mathrm{\mu s^{-1}}$, emitting a positron preferentially along the instantaneous spin
 direction at the time of decay.  Time spectra were constructed by simultaneously measuring the
 asymmetry of positron counts along axes both parallel and perpendicular to $\hat{c}$.

The evolution of a single muon spin in the presence of a local field, $\vec{B_{loc}}$ is given by
\begin{equation}
S_{z}(t) = \mathrm{cos^{2}(\theta) + sin^{2}(\theta)cos}(\gamma_{\mu}B_{loc}t),
\label{eq:Pmu}
\end{equation}
where $\theta$ is the angle between the local field and initial spin direction, and $\gamma_{\mu}$=0.0852$\mu s^{-1} G^{-1}$. A $\mathrm{\mu SR}$ asymmetry spectrum is built from an ensemble of such muons, and represents a convolution of Eq.~\ref{eq:Pmu} with the local distribution of magnetic fields.

Our unusual choice for initial muon polarization direction was designed to avoid the exceptional case where $\mathrm{sin^{2}(\theta)}$=0 in Eq.~\ref{eq:Pmu}. Muons are known to lie in one of eight positions, related to each other through reflections in x=0, y=0 and z=0 for a unit cell centered on the copper ion\cite{hitti90}. Since the proposed magnetic order does not satisfy all of these reflections, we can say with certainty that the local field does not lie along the initial (low symmetry) spin direction for every muon site. Some muons will always precess in the presence of such order, which might not have been the case if the initial muon polarization lay along a high symmetry direction, like $\hat{c}$.

\begin{figure}[tpb]
\centering
\subfigure{
   \label{fig:theory}
   \includegraphics[width=0.75\columnwidth]{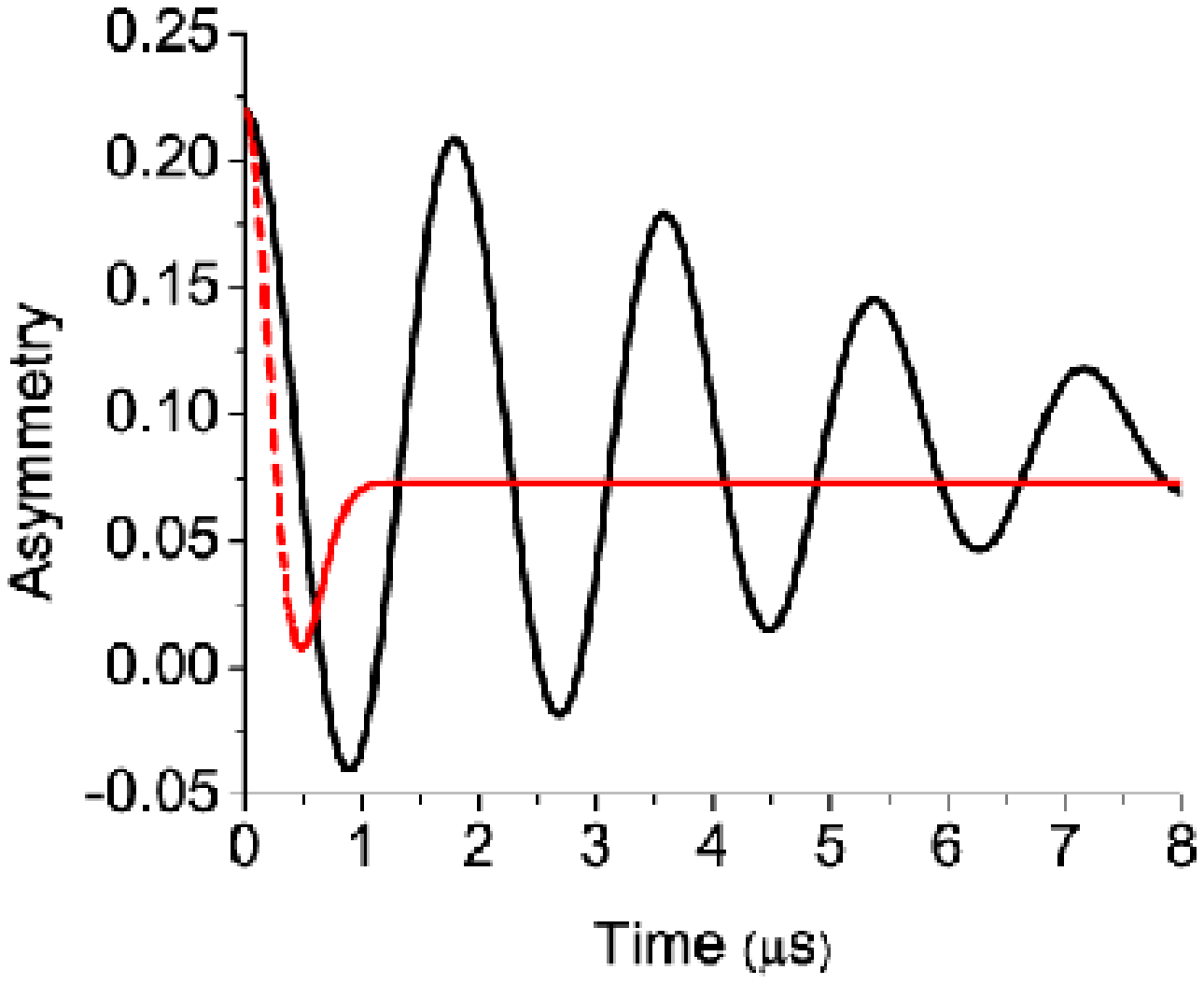}}
\subfigure{
   \label{fig:raw}
   \includegraphics[width=0.75\columnwidth]{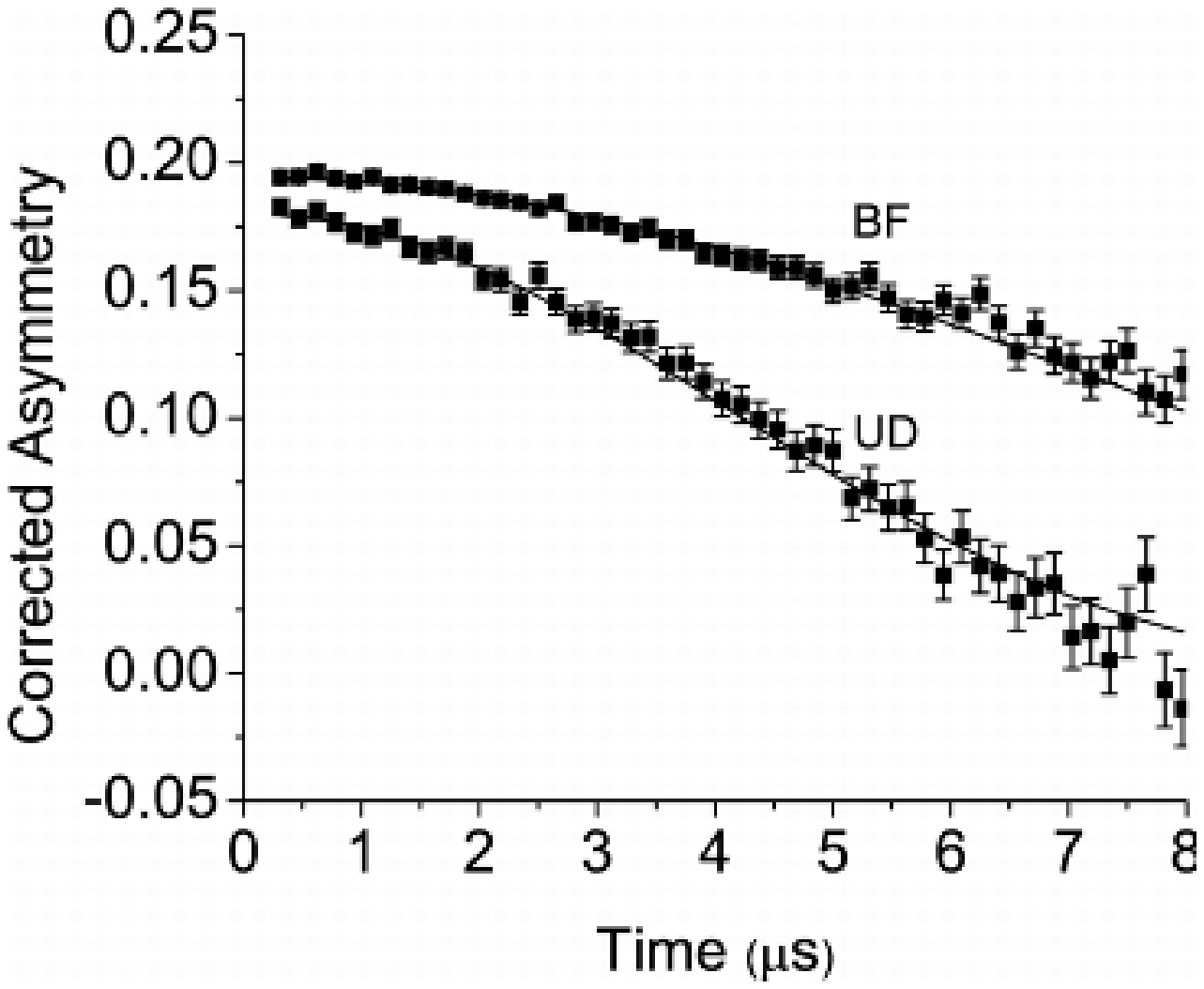}}
\caption{\label{fig:spectra}(Top) The expected $\mathrm{\mu SR}$ spectra for a material with the long-range current order seen in Fig.~\ref{fig:pattern} (black) or a distribution of fields as expected for a short-range-ordered state (red).  (Bottom) The actual $\mathrm{\mu SR}$ spectra observed in $\mathrm{La_{1.87}Sr_{0.13}CuO_{4}}$ at T=10K. Solid lines represent fits to Eq.~\ref{eq:KT}.}
\end{figure}

In a long-range ordered state, the distribution of local fields is very close to a delta-function, and $\mathrm{\mu SR}$ should see coherent precession at a frequency set by the average magnetic field at the muon site.
This is demonstrated by the black line in the top panel of Fig.~\ref{fig:spectra}, which was calculated (as described later)
 assuming the current pattern of
Fig.~1 with a moment of 0.05$\mathrm{\mu_{B}}$. Actual measured spectra from $\mathrm{La_{1.87}Sr_{0.13}CuO_{4}}$ at T=10K are displayed in the next panel. Despite this temperature being well below $T^{*}$ for this doping, in neither set of counters is there any sign of precession. Instead, the spectra are more indicative of a local field distribution dominated by randomly oriented nuclear dipole moments, which are orders of magnitude smaller than typical magnetic moments. The situation is the same at every temperature and in all three crystals.

Each spectrum was fit to the standard Kubo-Toyabe form expected for a Gaussian distribution of random fields:

\begin{equation}
G(t) = \dfrac{1}{3}+\dfrac{2}{3}(1-\Delta ^{2} t^{2})\mathrm{exp}(-\dfrac{1}{2} \Delta ^{2} t^{2}),
\label{eq:KT}
\end{equation}
where $\Delta / \gamma_{\mu}$ is the width of the magnetic field distribution. For small moments, this equation reduces to a Gaussian function, with $\Delta$ equal to the rate of relaxation.

\begin{figure}[tpb]
\includegraphics[width=0.75\columnwidth]{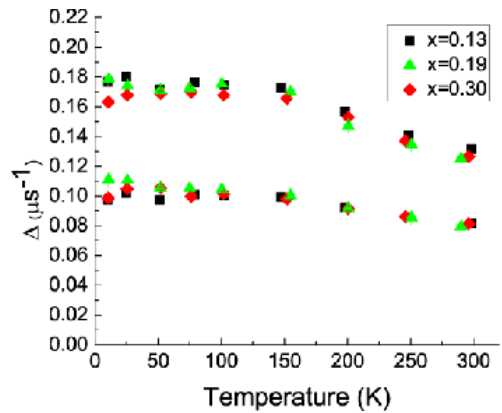}
\caption{\label{fig:rlx} The results of fits to Eq.~\ref{eq:KT}. Points on the upper (lower) curve represent fits from spectra with counters in (out of) the copper-oxide planes.}
\end{figure}

The results of these fits are summarized in Fig.~\ref{fig:rlx}. There is clearly no strong temperature dependence of the $\mathrm{\mu SR}$ spectra in any of the three crystals. At low temperatures, the rate of relaxation is constant, again consistent with what one would expect for a field distribution dominated by randomly oriented nuclear dipoles. There is a slight reduction in $\Delta$ above 175K in each of the crystals, but no dramatic change in the system at $\mathrm{T^{*}}$. In these materials, 175K marks the onset of muon diffusion\cite{sonier02}, the effect of which is to motionally narrow the dipolar relaxation; the slight decrease in $\Delta$ above this temperature reflects this effect.
The relaxation rate of the spectra from counters perpendicular to $\hat{c}$ were larger than those with counters parallel to $\hat{c}$ due to particulars of the distribution of dipoles in relation to the known muon site.

More important than the temperature dependence of the local field distribution is its dependence on Sr concentration. The gradual increase of pseudo-gap properties makes it difficult to correlate material properties via temperature. However, it is known that the pseudo-gap is strong in the underdoped materials and absent by x=0.30. Consequently, the observation that the three crystals investigated here have identical $\mathrm{\mu SR}$ spectra at every temperature puts tight restrictions on any TRSB fields which might be associated with the pseudo-gap state. Specifically, if one says that the average scatter of points in Fig.~\ref{fig:rlx} is 0.01$\mu s^{-1}$, an overestimation to be sure, then our data puts an upper limit of $\sim$0.2G on such fields at the muon site.

This is a significant restriction. By modeling the predicted current pattern by a series of infinitesimally thin wires,
we estimate that the expected local field at the known muon site\cite{hitti90} is approximately 41G for the $\it{smallest}$ reported moment in the pseudo-gap of YBCO\cite{fauque06}. This is more than two orders of magnitude larger than the upper bound provided by our measurements, and the difference is much greater than can be reconciled by more accurately modeling the orbital order.

Moreover, it is difficult to conceive of a mechanism by which $\mathrm{\mu SR}$ might miss such a large effect. As discussed earlier, symmetry considerations and the simultaneous use of two sets of orthogonal counters
with an initial 45$^\circ$ spin rotation preclude the possibility that we placed the initial muon spin direction parallel to the local field direction at every muon site. Nor does the exact muon site much matter. A map of the local fields throughout the entire unit cell shows comparably sized fields nearly everywhere in the presence of the proposed order.

The observed spectra are also inconsistent with short-range-order. It is well-known that LSCO has significant cation disorder, which in principle could destroy long-range current order in favor of a state with many small domains. However, even in the most extreme case where such disorder reduces the average magnetic field to zero, the local field environment would still be dominated a distribution of fields of similar size. This would manifest itself as a fast depolarization of the $\mathrm{\mu SR}$ asymmetry function, which would be indistinguishable from a long-range-ordered state at early times. This is demonstrated in Fig.~\ref{fig:spectra}, in which we have plotted (red) the expected asymmetry spectrum for a Gaussian distribution of local fields with a half-width of 41G.

There is a real concern that the presence of the poorly screened muon charge might perturb the local system
and destroy the current ordered state. Shekhter et al.\cite{shekhter08} estimate, based on correlation effects, that the local order should be suppressed within a radius of $\sim$3~unit cells in the nearest two planes. Specific modeling by us suggests that one would still be left with a local field of 0.6G at the muon site, three times our experimental limit, for the \textit{smallest} moment reported by neutron scattering. Thus, it is our opinion that muon perturbation does not reconcile our null result with theory. On this topic, we further note that STM has long observed poorly screened atomic ions directly beneath the copper-oxide planes of Bi2212\cite{STM}, and a corresponding suppression of superconductivity but not the pseudo-gap at these sites. Thus, the idea of a fragile pseudo-gap state which is strongly affected by electrostatic
charges does not seem to be consistent with what is currently known about these systems.

Sonier et al. have reported seeing small TRSB fields in the pseudo-gap state of two YBCO crystals with $\mathrm{\mu SR}$\cite{sonier01}. Recent polar Kerr effect measurements see the onset of weak ferromagnetism at similar temperatures\cite{xia07}, implying a common mechanism is at work. However, the latter observation is inconsistent with the proposed circulating current pattern, which has no net moment. In addition, both papers comment on the persistence of weak magnetism well above the pseudo-gap temperature.

The fields seen by Sonier are comparable to the size of the error bars in the present study. Thus, we cannot comment on the presence or absence of a similarly sized fields in LSCO, and this remains an open question. However, in neither case is the $\mathrm{\mu SR}$ data consistent with the current ordered state inferred from neutron scattering and ARPES measurements. Although the present study examines a different material and doping range than previous experiments, if such current order is a feature of a universal theory of the cuprates, it should be seen here.

It is important to note that the $\mathrm{\mu SR}$ results are not inconsistent with the more general conclusion that the pseudo-gap terminates at a QCP inside the superconducting dome. Indeed, for YBCO at least, a preponderance of evidence strongly supports such a conclusion\cite{sonier01,fauque06,xia07,mook08}. $\mathrm{\mu SR}$ does put tight restrictions on the form of possible pseudo-gap order parameters for LSCO. Yet, several QCP scenarios are still viable. For example, current patterns which have reflection symmetry across the Cu-O-Cu bonds would have a greatly reduced local field at the known muon site and might reconcile the $\mathrm{\mu SR}$ and neutron scattering results. Examples include the $\Theta_{I}$ state of Varma\cite{varma} and the DDW state of Chakravarty\cite{chakravarty01}.

One possible route to reconcile the absence of an effect in our measurement would be if the current pattern fluctuated between different equivalent arrangements. If the fluctuations were sufficiently rapid, the local field distribution would be motionally narrowed. With a static Gaussian distribution of 41G fields, a fluctuation rate larger than 200MHz would lower the relaxation rate below our detection limit. Such a fluctuation rate would still look static on neutron scattering timescales.

It is also conceivable that slight differences in crystal structure and material parameters in different families of superconductors stabilize different current patterns. Alternatively, the magnetic order seen by neutron scattering may be secondary to an alternate pseudo-gap order parameter, and only appear in a certain subset of cuprates materials. Such questions can only be answered by further studying different materials with multiple techniques. In the future, we plan to perform similar measurements on crystals where magnetic neutron scattering has already been observed.

The present results show that any time reversal symmetry breaking fields for La$_{2-x}$Sr$_x$CuO$_4$ must be less than about 0.2~G. Any universal theory of the cuprates must takes this into account.

This work was supported by NSERC and NSF-DMR-05-02706.  GJM and AAA are supported by NSERC Canada Graduate Scholarships.  We
appreciate the hospitality and technical assistance of the TRIUMF Centre for Molecular and Materials Science where
these experiments were performed.

\end{document}